\documentclass[11pt,a4paper]{article}
\usepackage{jheppub_kim}
\usepackage{epsfig}
\usepackage{amsmath}
\usepackage{graphicx}
 \usepackage{graphics}
 \usepackage[scriptsize,nooneline,hang]{caption}
\usepackage[hang,nooneline,scriptsize]{subfigure}
 
\begin{document}

\title{Perturbed thermodynamics and thermodynamic geometry of a static black hole in $f (R)$ gravity}

\author[a,b,c,d]{Sudhaker Upadhyay,}
\author[e]{Saheb Soroushfar}
\author[f]{and Reza Saffari}
\affiliation[a] {Department of Physics, K.L.S. College,  Nawada, Bihar  805110, India.}
\affiliation[b] {Department of Physics, Magadh University, Bodh Gaya, Bihar 824234, India}
\affiliation[c]{Visiting Associate, Inter-University Centre for Astronomy and Astrophysics (IUCAA) Pune, Maharashtra-411007}
\affiliation[d]{School of Physics, Damghan University, P. O. Box 3671641167, Damghan, Iran}
 \affiliation[e]{Faculty of Technology and Mining, Yasouj University, Choram 75761-59836, Iran}
\affiliation[f]{Department of Physics, University of Guilan, 41335-1914, Rasht, Iran}

\emailAdd{sudhakerupadhyay@gmail.com}
\emailAdd{soroush@yu.ac.ir}
\emailAdd{rsk@guilan.ac.ir}

\abstract
{In this paper, we consider a static black hole in $f(R)$ gravity. 
We recapitulate the expression for corrected thermodynamic entropy  
of this black hole due to small fluctuations around equilibrium.
Also, we study the geometrothermodynamics  (GTD)  of this black hole and investigate
the adaptability of the curvature scalar of geothermodynamic methods with 
phase transition points of this black hole. 
Moreover, we study the effect of correction parameter on thermodynamic behaviour  of this black hole.
We observe that  the singular point of the curvature scalar of 
Ruppeiner metric  coincides completely  with 
zero point of the heat capacity and the deviation occurs with  increasing correction parameter.}

\maketitle

\section{Introduction}  
The thermodynamics of black holes is a subject of interest these days.
Thermodynamic quantities of black holes such as  temperature, entropy and heat capacity;  
  thermodynamic properties such as phase transitions and thermal stability;  geometrical quantities 
such as horizon area and surface gravity;  thermodynamical variable such as cosmological 
constant have been studied in several papers ~\cite{Bekenstein,Hawking,Davies,Page,Cai,Kothawala,Myung,Eune,Gibbons,Dolan,Banerjee,Kubiznak,
Cao,Jahani,Chen,Mo,Zou,Xu,Frassino,Johnson,Dolan2,Xu2,Hendi,Mandal,Hendi2}.
Geometrical thermodynamics is other applied and important method to study phase transition 
of black holes. For present different implications of geometry in to usual thermodynamics, 
many endeavour are presented in various articles~\cite{Hermann,Weinhold,Ruppeiner,Salamon,Ruppeiner2,Quevedo,H. Quevedo}. 
It is well-known that  the larger black holes  in comparison to  the Planck scale 
have entropy proportional to its horizon area, so it is important to investigate the 
form of entropy  as one reduces the size of the black hole~\cite{Strominger,Baez,Carlip,Solodukhin,Ashtekar,Das}. 
It is found that the area-law of entropy of a thermodynamic
system  gets correction due to  the fluctuations around equilibrium \cite{Das}.
These corrections are evaluated through both 
microscopically  and using a stringy embedding of the Kerr/CFT correspondence
\cite{stro}.
The leading-order correction to the  entropy area-law   is also estimated 
through   the Cardy formula \cite{card}.
Various check-ups  confirm that 
the entropy   due to the  perturbation around thermal equilibrium of the black hole has logarithmic correction.

The study  of fluctuations  to the black holes thermodynamics is a subject 
of current interests. For instance,      the thermal fluctuations  corrects the thermodynamics of higher dimensional AdS black hole, where it has been found that   the Van der Waals black hole is completely stable in presence of the logarithmic correction \cite{behsud}.
The leading-order corrections to the Gibbs free energy, charge and total mass densities of charged quasitopological and charged rotating quasitopological black holes due to thermal fluctuation are derived, where the stability and bound points of such black holes under effect of leading-order corrections are also discussed
\cite{sudhaker}. The effect of thermal fluctuations on the thermodynamics of a black
geometry with hyperscaling violation is also studied recently \cite{lina}.
The  thermal corrections on the thermodynamics and stability  of   Schwarzschild-Beltrami-de Sitter black hole \cite{sud2}, non-rotating BTZ black
hole \cite{sudu}, charged rotating AdS black holes \cite{sudu1}, G\"{o}del black hole \cite{godel},
and   massive black hole 
 \cite{sud3} have been investigated recently.  
 Also, the effects of thermal fluatuation  on the Ho\v rava-Lifshitz black hole thermodynamics  and their stability  are    discussed \cite{sud1}.  The leading-order correction to modified Hayward black hole is  derived and found that correction term reduces the pressure and internal energy
of the Hayward black hole \cite{behn}.  

On the other hand, in order to introduce
 concepts of geometry into ordinary thermodynamics many efforts have been made.
 In this regard, the implication of thermodynamic
phase space as a differential manifold is defined, which embeds a special subspace of thermodynamic
equilibrium states \cite{Hermann}.
Many different perspective with different goals and reviews  in $f(R)$ gravity 
have been studied in the past decade until now \cite{Brans:1961sx,Riess:1998cb,Fujii:2003,Brax:2003fv,Buchdahl:1983zz,Starobinsky:1980te,Bamba:2008ja,Dastan:2016vhb,Soroushfar:2016esy,Chakraborty:2015bja}.  Keeping importance of $f(R)$ gravity in mind, 
our motivation is to study the effect of thermal fluctuation on the 
thermodynamics and thermodynamic geometry of a static black hole in $f (R)$ gravity.

In this paper, we obtain corrected thermodynamic entropy 
and investigate thermodynamic quantities and thermodynamic geometric methods  
for a black holes in $f(R)$ gravity. We find  that the Hawking temperature is a decreasing 
function of horizon radius.  In order to evaluate the correction to
 the entropy of a static black hole in $f(R)$ gravity due to   thermal fluctuation, we exploit the
 expressions of Hawking temperature and uncorrected specific heat. Moreover, utilizing standard thermodynamical relation, we compute first-order corrected mass and heat capacity.
 Here, we observe that the uncorrected 
mass of the black hole has a maximum value
at $r_{+}=r_{m}=2.31$, and takes zero value at two points, $r_{+}=0$ and $r_{+}=4$. 
For the corrected mass with positive correction coefficient $ \alpha $, 
the number of zero points of mass are not changing, but they appear with larger horizon radius.
 In   case of heat capacity, we find that  
 the uncorrected heat capacity is in the negative region (unstable phase) for $0<r_{+}<r_{m} $  and it takes type one phase transition    at  $ r_{+}=r_{m} $  for which $ C(r_{+}=r_{m})=0 $. For  $ r_{+}>r_{m} $, it
becomes positive valued (stable).
So, without considering thermal fluctuation, this black hole has a type one phase transition.
However, as long as we turn on effects of thermal fluctuation,
the number of zero points of the heat capacity changes to the three zero points.
Furthermore, we analyse the thermodynamic geometry of such black holes.  In order to do so, we 
plot  thermodynamic quantities and curvature scalar of Weinhold, Ruppeiner 
and GTD methods in terms of horizon radius. We find that
 the singular points of curvature scalar of Weinhold and GTD methods, 
for both with and without thermal fluctuations, are not coinciding with zero point of 
heat capacity (the phase transition points) which suggests that we are unable to get any physical information 
about the system with these two methods. 
However, without considering correction, we find that the heat capacity has only one zero, 
and the singular point of the curvature scalar of Ruppeiner metric is completely coinciding 
with it. The heat capacity under the effects of thermal fluctuation has 
three zero points  and not all the singular points of the curvature scalar of Ruppeiner metric are 
not completely coinciding with zero points  of such black hole but only one 
of the singular point of the curvature scalar of Ruppeiner metric is completely coincide with one 
of the zero point of the heat capacity.  This suggests that, by increasing $ \alpha $,  the above adaptation is reducing.

This paper is organized as follows,
in Sec.~\ref{section2}, we recapitulate the expression for corrected thermodynamic entropy  
of black hole due to small fluctuations around equilibrium, 
Then, in Sec.~\ref{section3}, we briefly review the static $f(R)$ black hole solution 
and its thermodynamics. Also, we study modified thermodynamics due to thermal fluctuations and 
thermodynamic geometry methods for this black hole. 
We conclude our results in Sec.~\ref{section4}.

\section{Entropy under thermal fluctuation}\label{section2}
In this section, we recapitulate the expression for corrected thermodynamic entropy  
of black holes due to small fluctuations around equilibrium.  To do so, let us  begin by
defining the density of states with fixed energy  as \cite{boh,rk}
\begin{eqnarray}
\rho(E) =\frac{1}{2\pi i}\int_{c-i\infty}^{c+i\infty}e^{{\cal S}(\beta)}d\beta,\label{rho}
\end{eqnarray}
where the exact entropy, ${\cal S}(\beta)=\log Z(\beta)+\beta E$,   depends on temperature $T(=\beta^{-1})$ explicitly. 
So, this (exact entropy)  is not just its value at equilibrium. 
The exact entropy corresponds to the  sum of entropies  of subsystems of the thermodynamical 
system. This  thermodynamical systems are small enough to be considered in equilibrium.
To investigate the form of exact entropy, we  
  solve the  complex integral (\ref{rho}) by considering the method of steepest descent around the saddle point $\beta_0 (={T_H^{-1}})$ such that $\left.\frac{\partial {\cal S}(\beta)}{\partial \beta}\right|_{\beta=\beta_0}=0$.  
Now, performing Taylor expansion of  exact entropy  around the saddle point $\beta=\beta_0$ 
leads to
\begin{eqnarray}
{\cal S}(\beta)={\cal S}_0+\frac{1}{2}(\beta-\beta_0)^2 \left(\frac{\partial^2{\cal S}(\beta)}{\partial \beta^2}\right)_{\beta=\beta_0}+ \mbox{(higher order terms)}.\label{s}
\end{eqnarray}
Here ${\cal S}_0(={\cal S}(\beta_0))$ is the leading-order entropy.
With this value of ${\cal S}(\beta)$,   the density of states (\ref{rho}
becomes
\begin{eqnarray}
\rho(E) =\frac{e^{{\cal S}_0}}{2\pi i}\int_{c-i\infty}^{c+i\infty}\exp\left[\frac{1}{2}(\beta -\beta_0)^2 \left(\frac{\partial^2{\cal S}(\beta)}{\partial \beta^2}\right)_{\beta=\beta_0}\right]d\beta.
\end{eqnarray}
This further leads to
\cite{Das}
\begin{eqnarray}
\rho(E)=\frac{e^{{\cal S}_0}}{\sqrt{2\pi \left(\frac{\partial^2 {\cal S}(\beta)}{\partial \beta^2}\right)_{\beta=\beta_0}}},
\end{eqnarray}
where  $c=\beta_0$ and $\left.\frac{\partial^2 {\cal S}(\beta)}{\partial \beta^2}\right|_{\beta=\beta_0}>0$  are chosen. 

Taking logarithm of above density of states yields 
the microcanonical entropy  density due to the small fluctuations around
thermal equilibrium 
\begin{eqnarray}
S={\cal S}_0-\frac{1}{2}\log \left(\frac{\partial^2 {\cal S}(\beta)}{\partial \beta^2}\right)_{\beta=\beta_0}+ \mbox{(sub-leading terms)}.
\end{eqnarray}
One can determine the form of   $\left.\frac{\partial^2{\cal S}(\beta)}{\partial \beta^2}\right|_{\beta=\beta_0}$ by considering the most general form  of the exact entropy  density  ${\cal S}(\beta)$. Das et al. in \cite{Das} have found  the form of   $\left.\frac{\partial^2{\cal S}(\beta)}{\partial \beta^2}\right|_{\beta=\beta_0}=C_0T_H^2$,  which leads to 
the leading-order corrected entropy 
as
\begin{equation}\label{corS}
S = {\cal S}_{0} -\frac{1}{2} \ln (C_{0} T_H^{2}),
\end{equation}
where $C_{0}=\left(\frac{\partial E}{\partial T}\right)_{T_H}$ is specific heat and $T_H$ is Hawking temperature.
Now,   a general
expression for the corrected entropy can be written  as following:
 \begin{equation} 
S = {\cal S}_{0} -\alpha \ln (C_{0} T_H^{2}).\label{cor}
\end{equation}
Here we introduced a parameter $\alpha$ by hand to track corrected terms where $
\alpha$ can have two values either $0$ or $0.5$. In case of $\alpha =0 $ we have original results  and  for $\alpha =0.5$ we have corrected entropy. 
Now, we would like to study the
effects of such correction term on the  thermodynamic geometry of black
holes in $f(R)$ gravity.

\section{Corrected thermal behaviour of a static black hole in $f(R)$ gravity}\label{section3}
 In this section we briefly review the static $f(R)$ black hole solution and its thermodynamics. 
The generic form of the action (in unit $G = c = \hbar = \kappa = 1$) is given by
\begin{equation}\label{action}
I=\frac{1}{2}\int d^{4}{x}\sqrt{-g}f(R)+S_{mat},
\end{equation}
where $S_{mat}$ refers to the matter part of the action and   expression for $f(R)$ gravity is considered as
\begin{eqnarray}
f(R) =R+\Lambda +\frac{R+\Lambda}{R/R_0+2/\eta}\ln\frac{R+\Lambda}{C}.
\end{eqnarray}
Here $\Lambda$ is the cosmological constant, $C$ is integration constant and  $R_0=6\eta^2/d^2$ with    free parameters $\eta$ and $d$. 
The spherically symmetric solution of the field equations of the action~(\ref{action}) is
\begin{equation}\label{ds2}
ds^{2}=-B(r)dt^{2}+B(r)^{-1}dr^{2}+r^{2}(d\theta^{2}+\sin^{2}\theta d\varphi^{2}),\quad
\end{equation}
where
\begin{equation}\label{B(r)}
B(r)=1-\frac{2m}{r}+\beta_{1} r-\frac{1}{3}\Lambda{r}^{2},
\end{equation}
here  parameter $m$ is related to the mass of the black hole, $\Lambda$ is the cosmological 
constant and $\beta_{1}$  is a real constant~\cite{Saffari:2007zt,Soroushfar:2015wqa}.
Eqs. (\ref{ds2}) and (\ref{B(r)}), for $ \beta_1=0 $,  convert to the space time of Schwarzschild AdS black hole  and for  $ \beta_{1}=0$ and $\Lambda=0 $,  represent the Schwarzschild black hole. 
If $ r_{+} $ denotes the radius of the event horizon, by setting $ B(r)=0 $,
the mass of the black hole, using the relation between entropy ${\cal S}_{0}$  and event horizon radius
$r_{+}$, $ ({\cal S}_{0}=\pi r^{2}_{+})$, is given by
\begin{equation}\label{mass}
{M}_0({\cal S}_{0},l,\beta)=\frac{l^2\pi^{1/2}\beta_1{\cal S}_{0}+ l^2\pi {\cal S}_{0}^{1/2}- {\cal S}_{0}^{3/2}}{2l^2\pi^{3/2}},
\end{equation}
where parameter $ l $  is related to the cosmological constant $\Lambda$  as follows
~\cite{Tharanath:2014ika},
\begin{equation}\label{Lambda}
\Lambda=\frac{3}{l^{2}}.
\end{equation}
The Hawking temperature  ($T_H=\frac{\partial { M_0}}{\partial {\cal S}_{0}}$) and heat capacity  ($C_0=T_H\frac{\partial   {\cal S}_{0} }{\partial T_H}$) are calculated as \cite{re}
\begin{eqnarray}\label{T0}
T_H&=&\frac{2\beta_1 l^2\pi^{1/2}{\cal S}_{0}^{1/2}-3{\cal S}_{0}+l^2\pi}{4\pi^{3/2}l^2{\cal S}_{0}^{1/2}},
\end{eqnarray}
\begin{eqnarray}\label{C0}
C_0&=&-\frac{4\beta_1l^2\pi^{1/2}{\cal S}_{0}^{3/2}-6{\cal S}_{0}^{2}+2l^2\pi{\cal S}_{0}}{l^2\pi+3{\cal S}_{0}}.
\end{eqnarray}
Eqs. (\ref{mass}), (\ref{T0}) and (\ref{C0}) for $ \beta_1=0 $,  convert to 
 \begin{eqnarray}\label{M-sads}
 M_{(Schwarzschild-AdS)}&=&1/2\,{\frac {\sqrt {S} \left( \pi \,{l}^{2}-S \right) }{{\pi }^{3/2}{l
 		}^{2}}} ,
 \end{eqnarray}
 \begin{eqnarray}\label{T-sads}
T_{(Schwarzschild-AdS)}&=&1/4\,{\frac {\pi \,{l}^{2}-3\,S}{{\pi }^{3/2}\sqrt {S}{l}^{2}}} ,
\end{eqnarray}
 \begin{eqnarray}\label{C-sads}
C_{(Schwarzschild-AdS)}&=&-2\,{\frac { \left( \pi \,{l}^{2}-3\,S \right) S}{\pi \,{l}^{2}+3\,S}}   ,
\end{eqnarray}
and for  $ \beta_{1}=0$ and $\Lambda=0 $, convert to

\begin{eqnarray}\label{M-s}
M_{(Schwarzschild)}&=&1/2\,{\frac {\sqrt {S}}{\sqrt {\pi }}} ,
\end{eqnarray}
\begin{eqnarray}\label{T-s}
T_{Schwarzschild}&=&1/4\,{\frac {1}{\sqrt {\pi }\sqrt {S}}}   ,
\end{eqnarray}
\begin{eqnarray}\label{C-s}
C_{Schwarzschild}&=&-2\,S   ,
\end{eqnarray}
which represent the Schwarzschild-AdS and Schwarzschild black holes respectively. 

\begin{figure}[h]
	\centering	
	\includegraphics[width=0.4\textwidth]{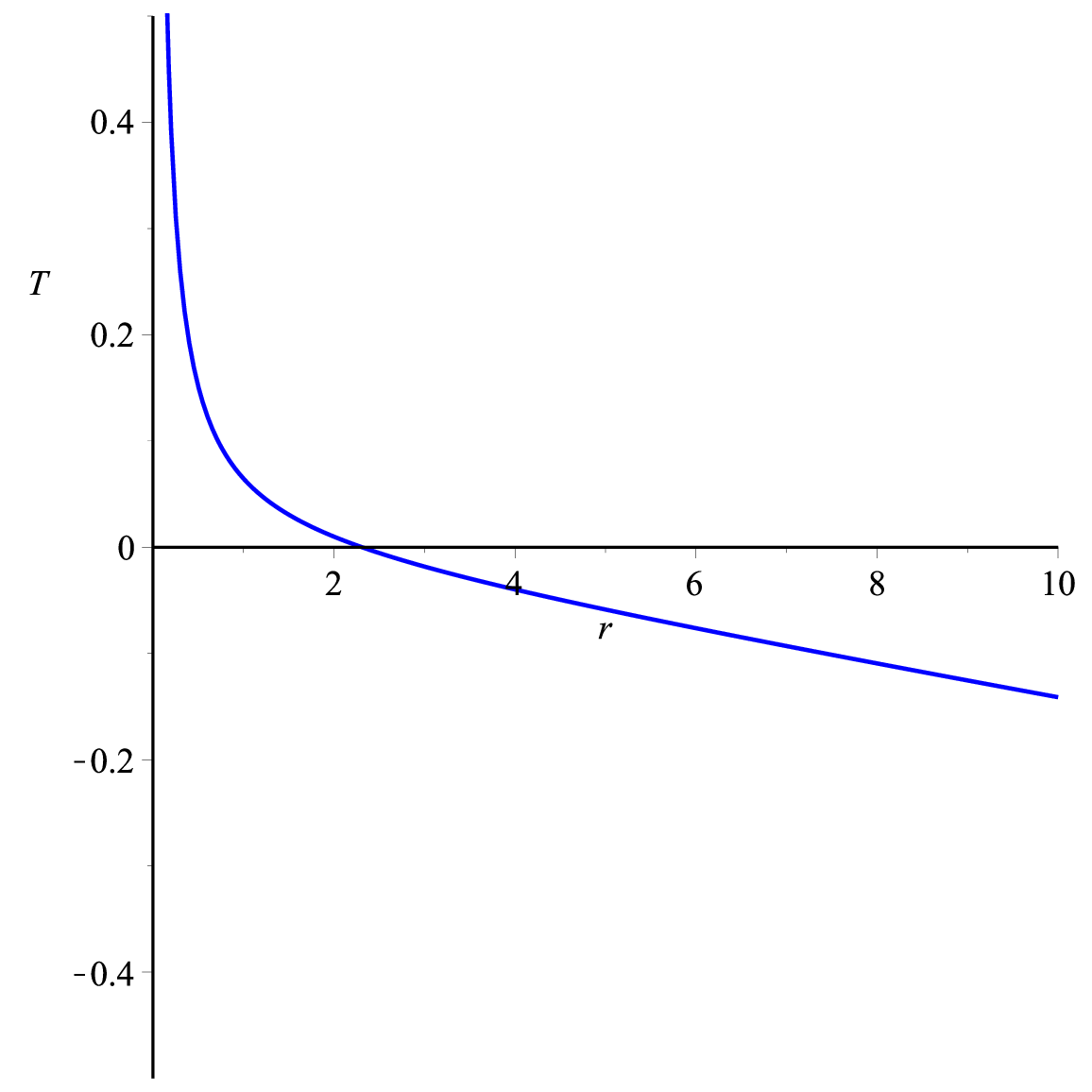}	
	\caption{Variation of temprature of the black hole in terms of horizon radius $ r_{+} $. $ l =4.0 $, $ \beta_{1} =10^{-4} $.}
	\label{pic:staticT}
\end{figure}

\begin{figure}[h]
	\centering	
	\includegraphics[width=0.4\textwidth]{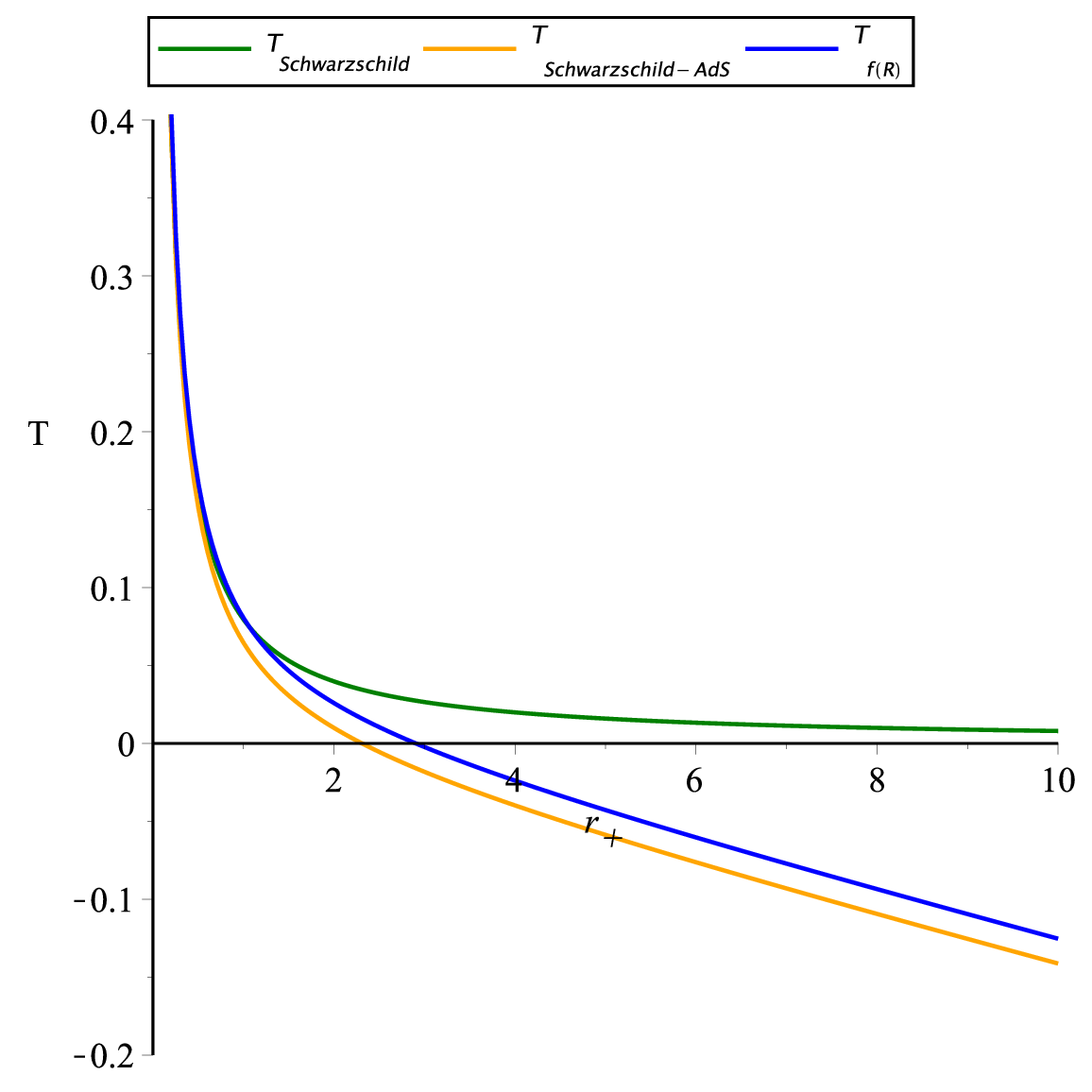}	
	\caption{Comparison of variation of temprature in terms of horizon radius $ r_{+} $. $ l =4.0 $, $ \beta_{1} =10^{-4} $, for $ f(R) $ black hole, $ l =4.0 $, $ \beta_{1} =0 $, for Schwarzschild-AdS black hole, and $ l =0 $, $ \beta_{1} =0 $, for Schwarzschild black hole.}
	\label{pic:staticNT}
\end{figure}

We shall utilize these expressions  of Hawking temperature  and heat capacity  to compute the
corrected entropy due to thermal fluctuations.
The  behavior of Hawking temperature in terms of horizon radius $r_{+}$ can be seen   in     Fig. \ref{pic:staticT}.  
From this plot, it is obvious   that the
temperature is a decreasing function of $r_{+}$ and takes positive value only in a
particular range of $r_{+}$, then at $r_{+}=r_{m}$, it reaches into zero, 
and after that, it falls into negative region.
Also, Fig. \ref{pic:staticNT} shows the comparison of variation of temprature in terms of horizon radius for a static black hole in $ f(R) $ gravity and  the Schwarzschild-AdS and Schwarzschild black holes of standard general relativity.
As we can see, the temperature behavior for a static black hole in $f(R)$ gravity and the Schwarzschild-AdS black hole is somewhat the same, i.e it is initially positive and becomes negative as the event horizon increases. But for the Schwarzschild black hole, the temperature  is always positive. 

\subsection{Modified thermodynamics due to thermal fluctuations}

\begin{figure}[h]
	\centering
	 \subfigure[]{
		\includegraphics[width=0.4\textwidth]{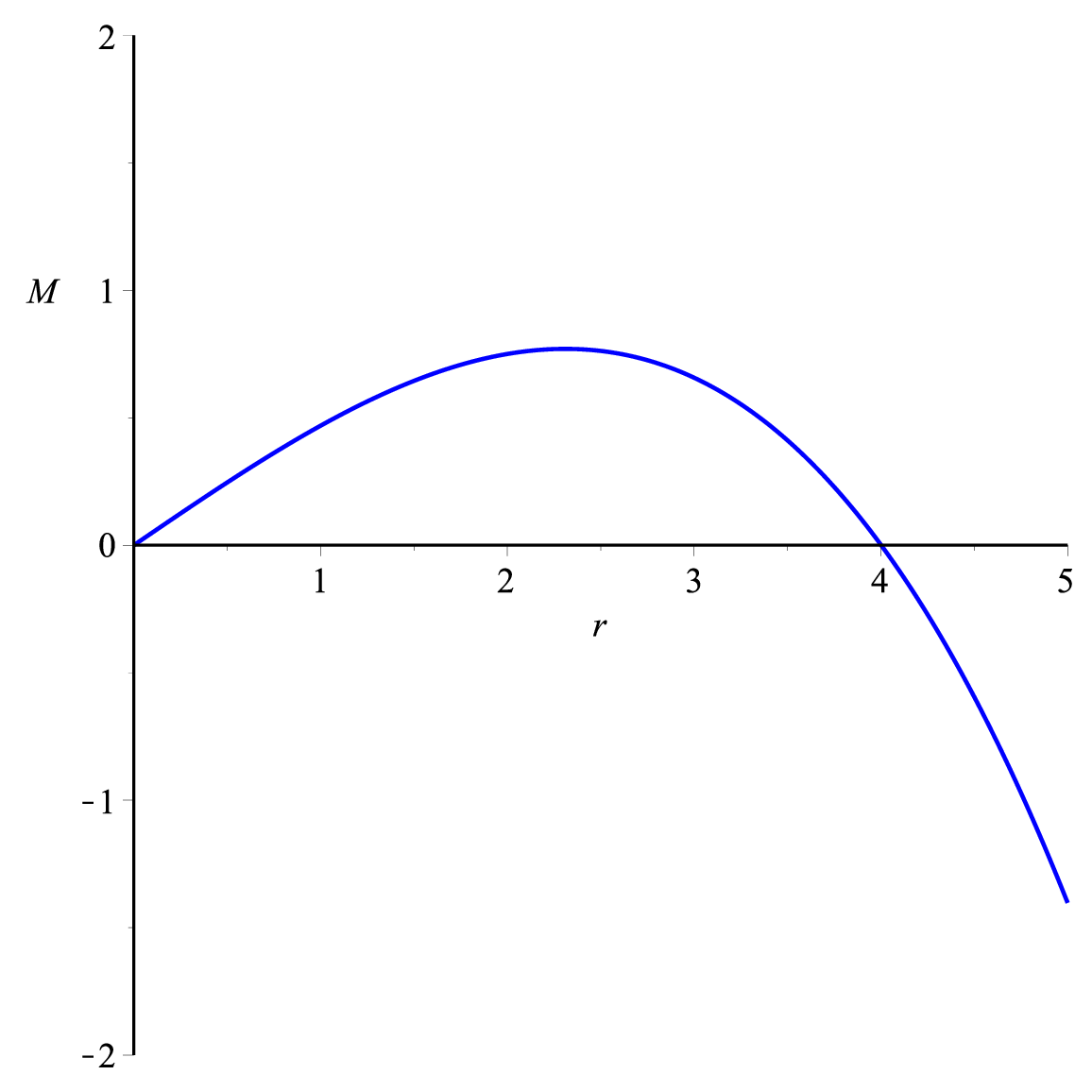}
	}
	\subfigure[]{
		\includegraphics[width=0.4\textwidth]{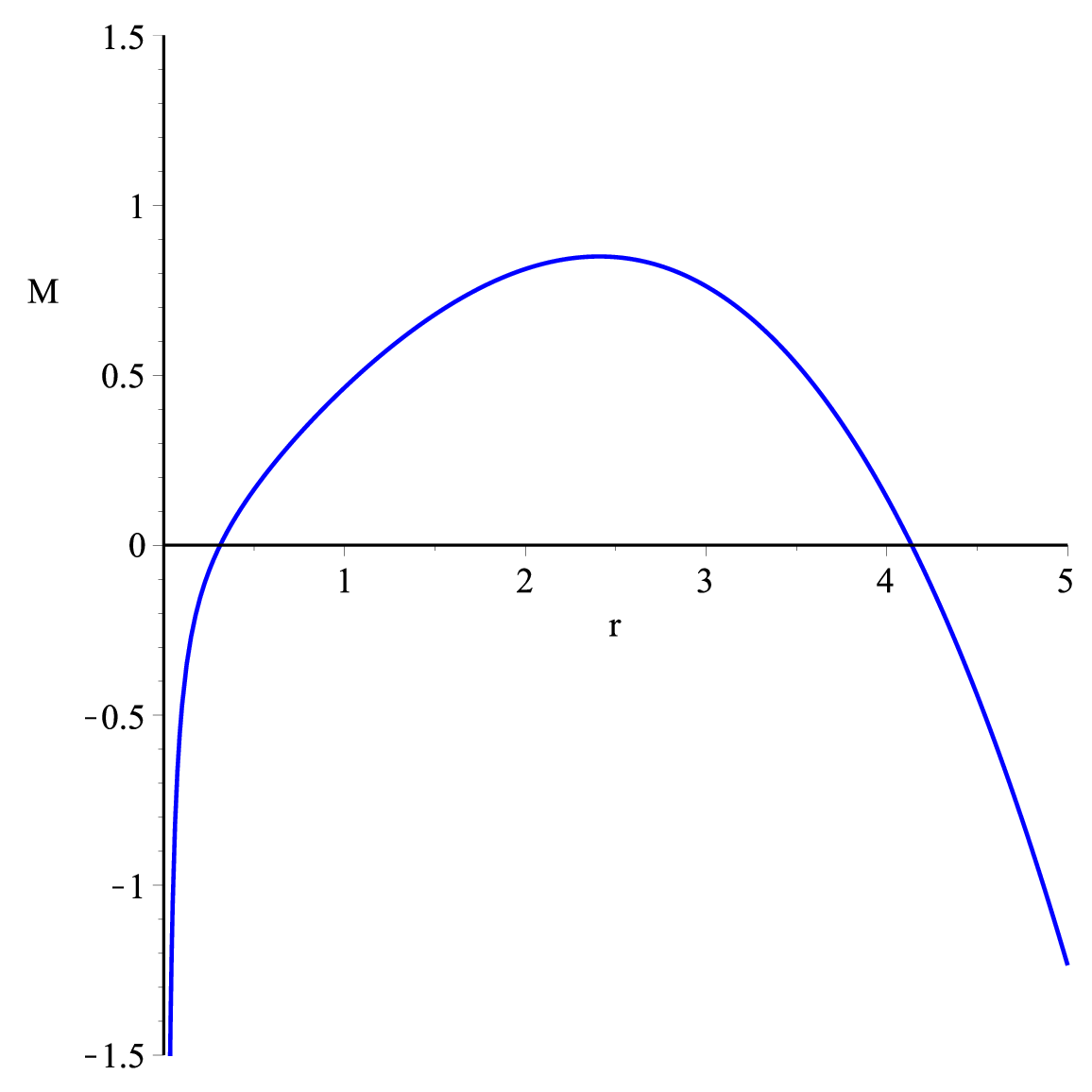}
	}
	\caption{Variation of mass in terms of horizon radius of a static black hole $ r_{+} $ for $ l=4.0 $, $ \beta_{1} =10^{-2} $ and $ \alpha=0 $, $ \alpha=0.5 $ for (a) and (b), respectively.}
 \label{pic:static-M}
\end{figure}

\begin{figure}[h]
	\centering
	\subfigure[]{
		\includegraphics[width=0.4\textwidth]{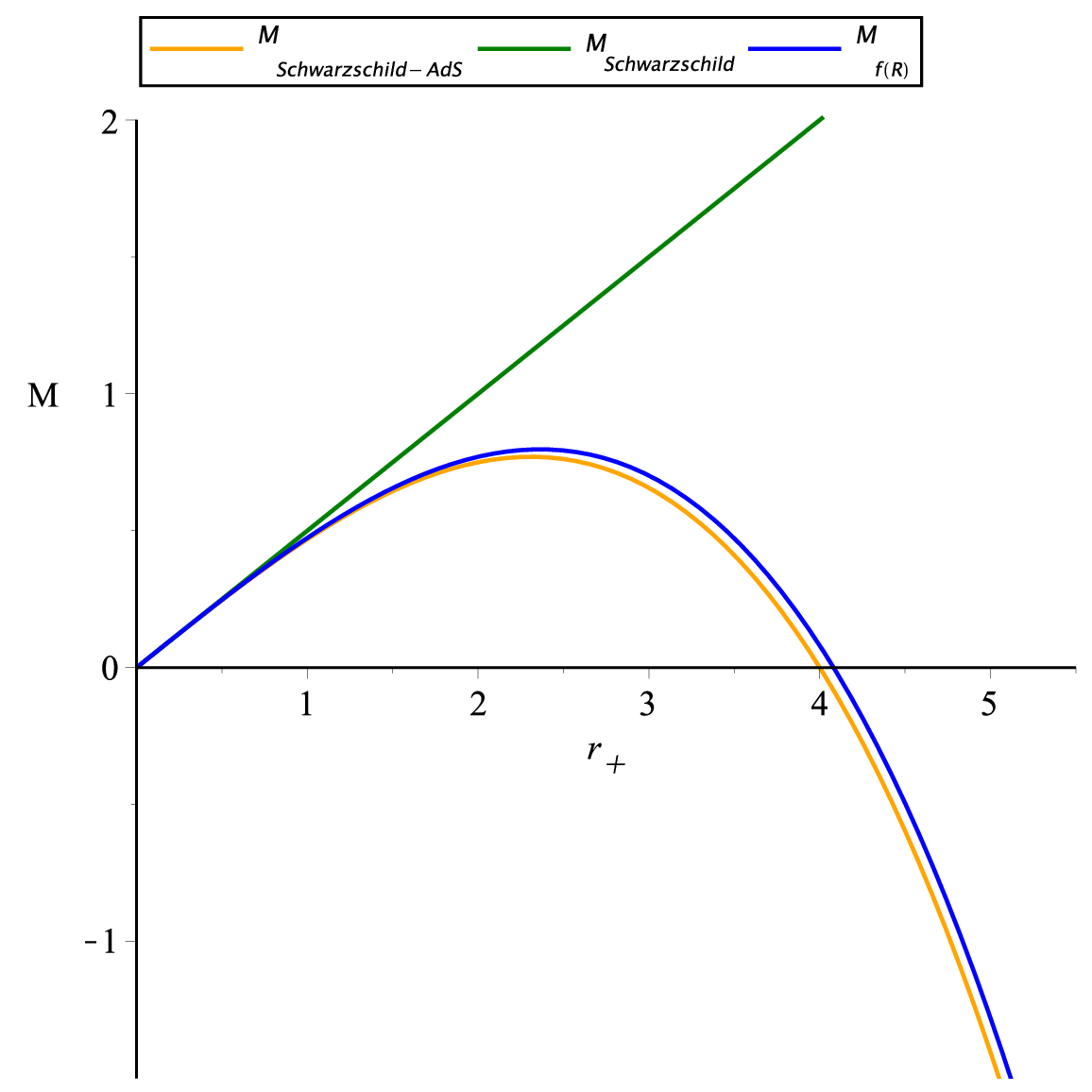}
	}
	\subfigure[]{
		\includegraphics[width=0.4\textwidth]{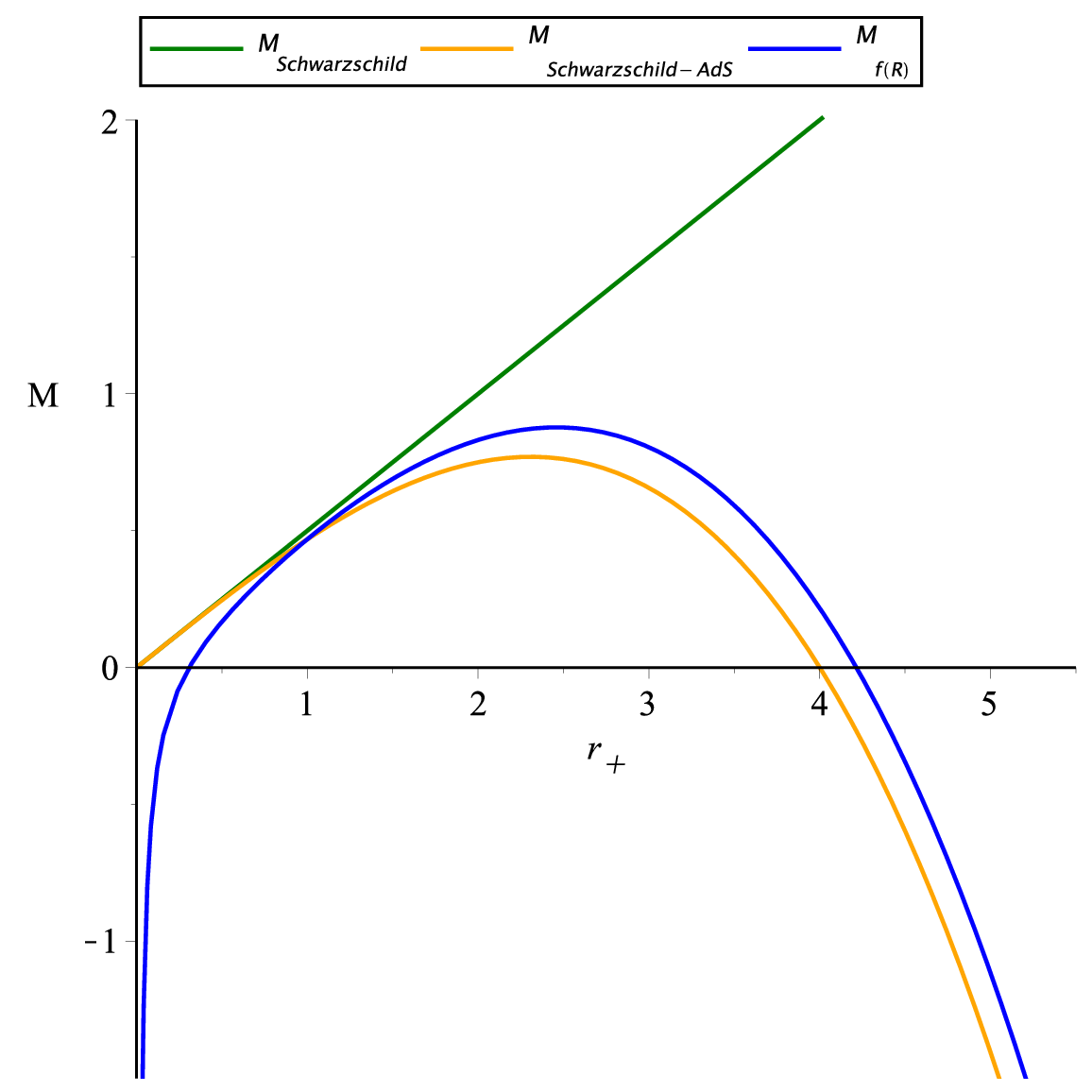}
	}
	\caption{Comparison of variation of mass in terms of horizon radius $ r_{+} $. $ l =4.0 $, $ \beta_{1} =10^{-2} $, for $ f(R) $ black hole, $ l =4.0 $, $ \beta_{1} =0 $, for Schwarzschild-AdS black hole, and $ l =0 $, $ \beta_{1} =0 $, for Schwarzschild black hole. Also, $ \alpha=0 $, $ \alpha=0.5 $ for (a) and (b), respectively.}
	\label{pic:staticN-M}
\end{figure}

\begin{figure}[h]
	\centering
	 \subfigure[]{
		\includegraphics[width=0.4\textwidth]{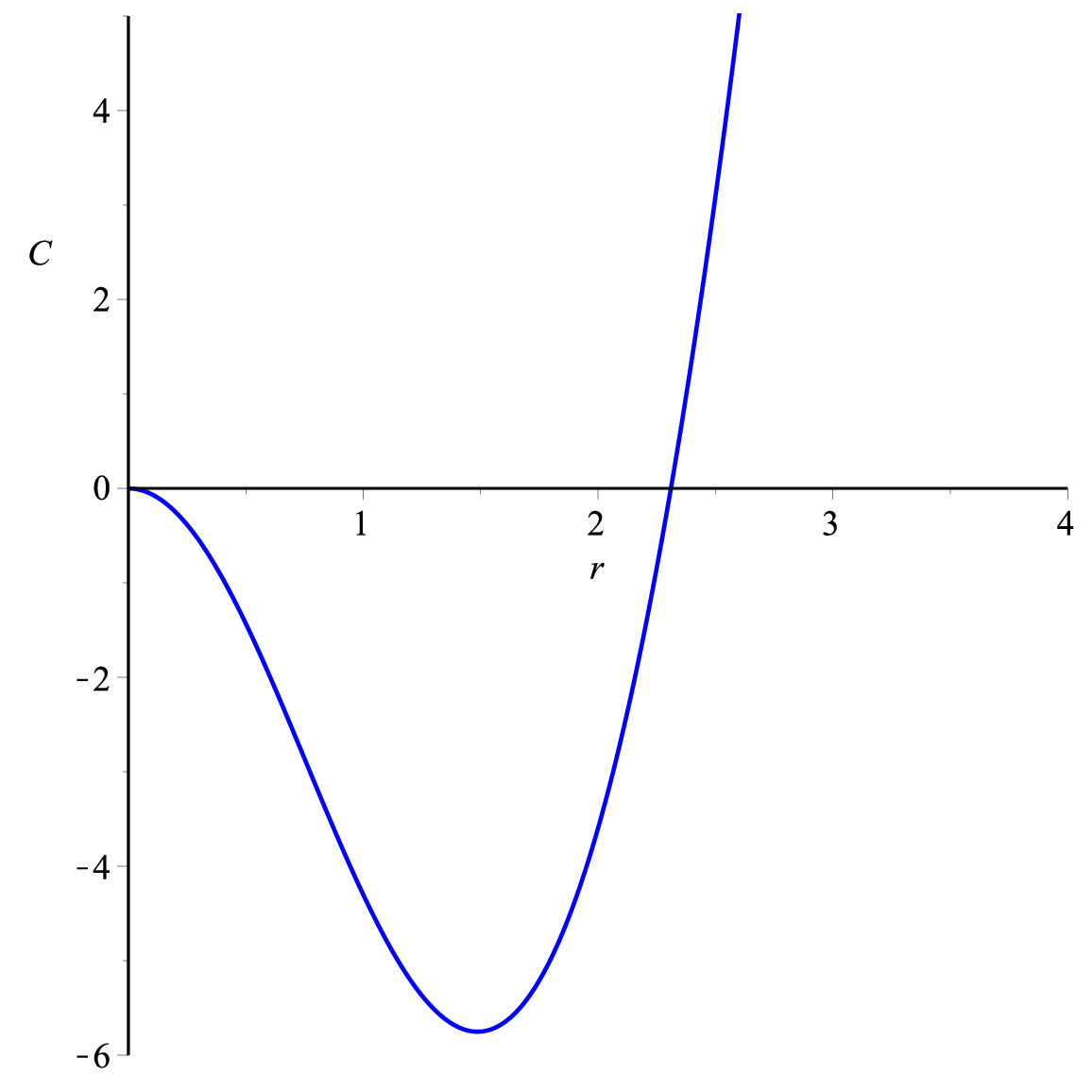}
	}
	\subfigure[]{
		\includegraphics[width=0.4\textwidth]{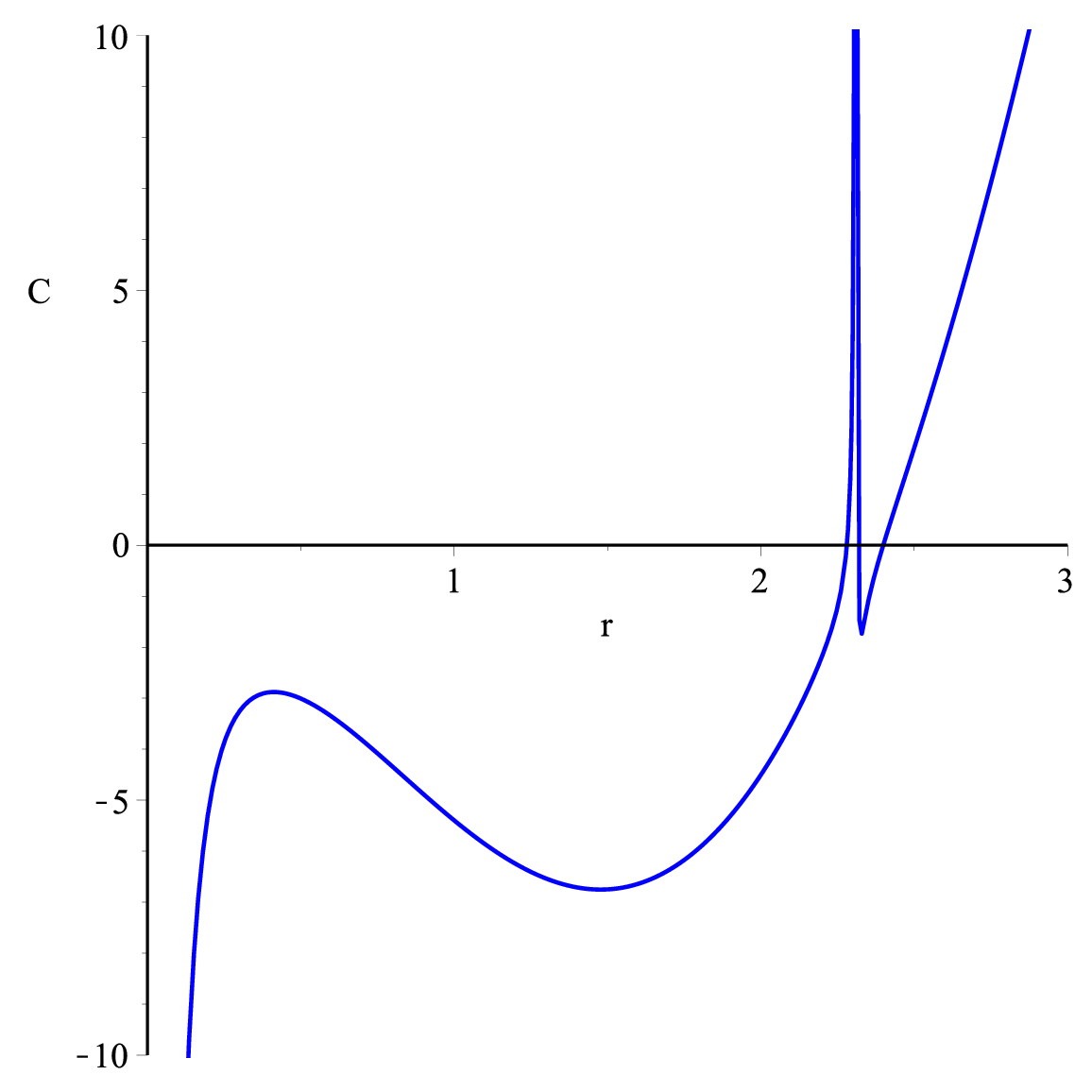}
	}
	\caption{Variation of heat capacity in terms of horizon radius of a static black hole $ r_{+} $ for $ l=4.0 $, $ \beta_{1} =10^{-4} $ and $ \alpha=0 $, $ \alpha=0.5 $ for (a) and (b), respectively.}
 \label{pic:static-C}
\end{figure}

\begin{figure}[h]
	\centering
	\subfigure[]{
		\includegraphics[width=0.4\textwidth]{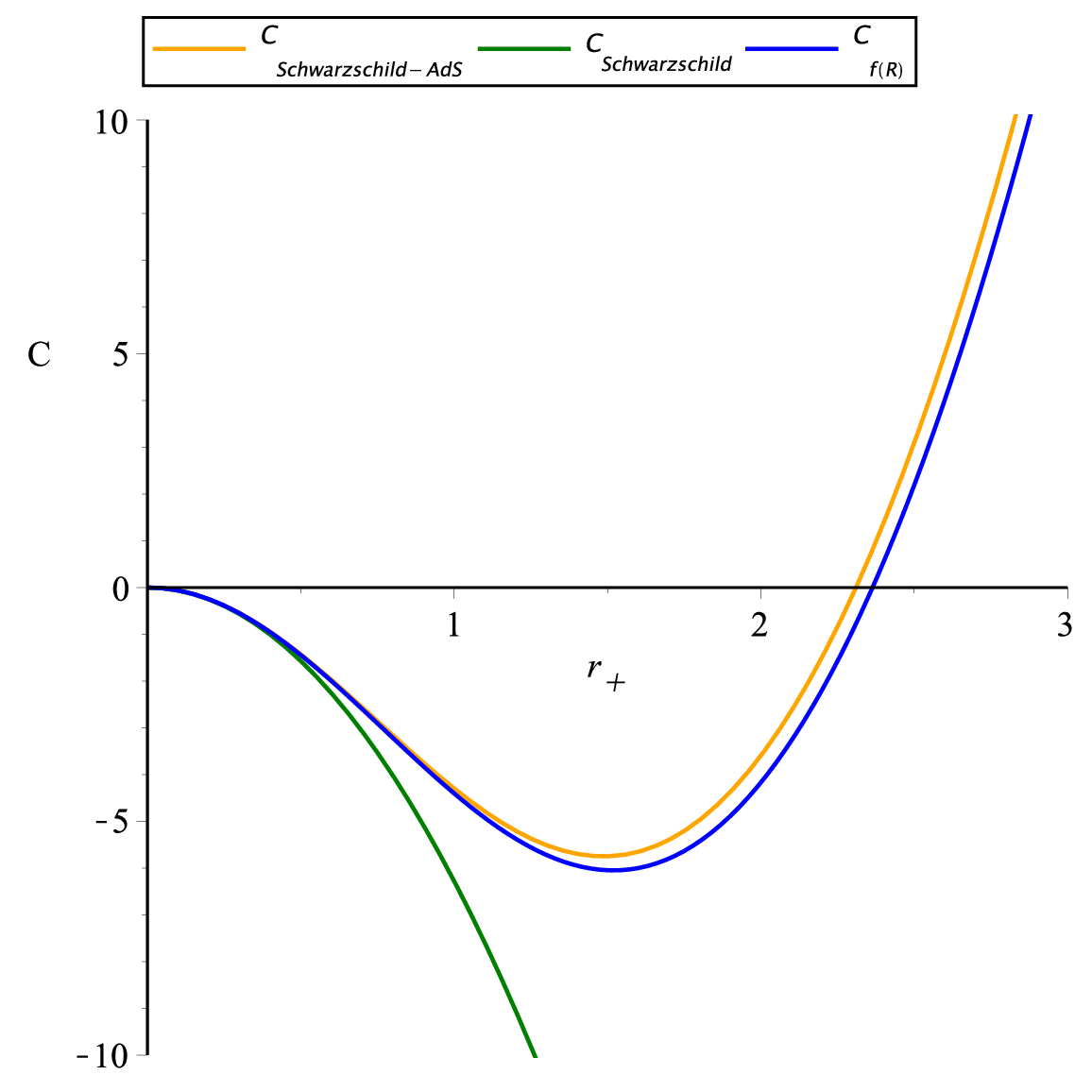}
	}
	\subfigure[]{
		\includegraphics[width=0.4\textwidth]{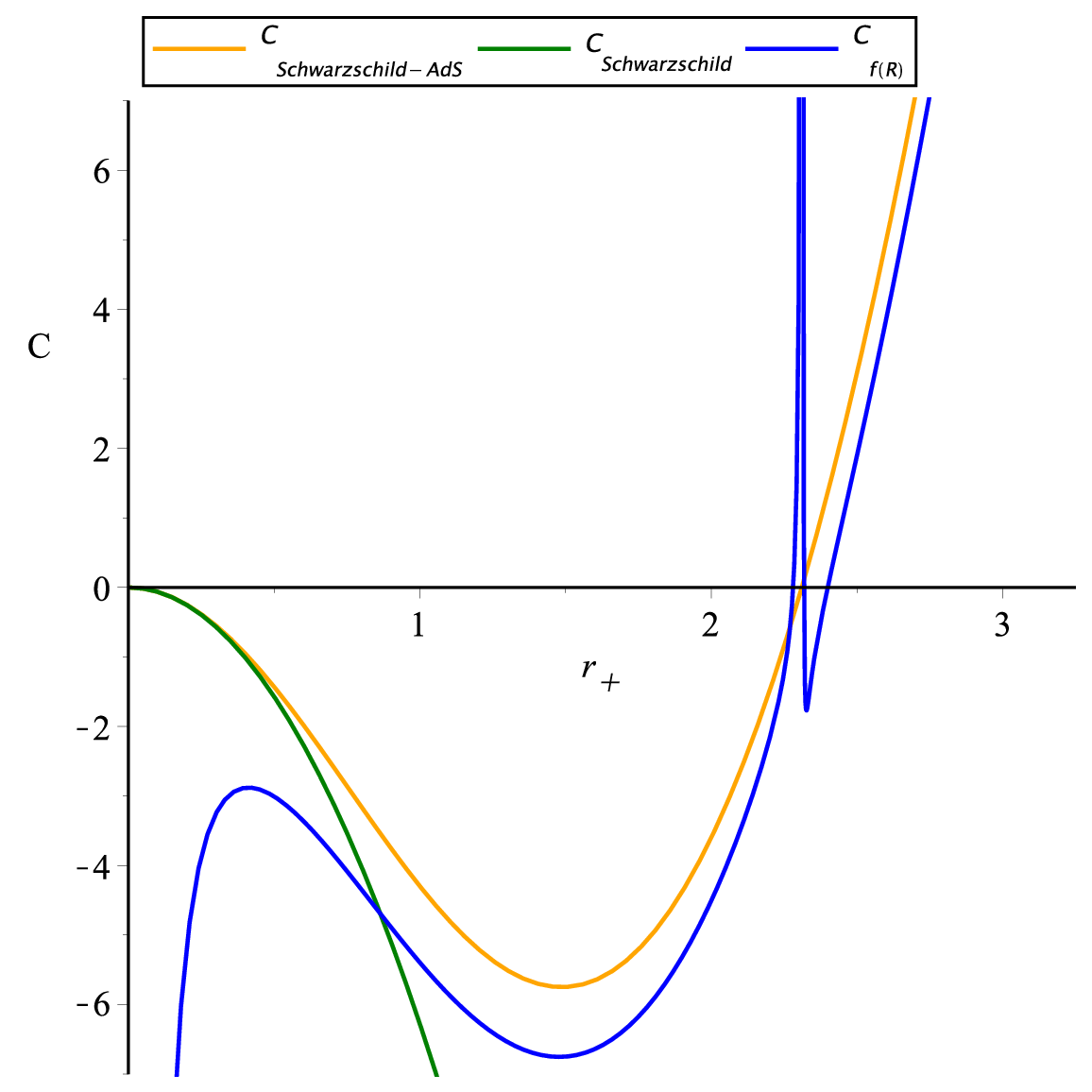}
	}
	\caption{Comparison of variation of heat capacity in terms of horizon radius $ r_{+} $. $ l =4.0 $, $ \beta_{1} =10^{-4} $, for $ f(R) $ black hole, $ l =4.0 $, $ \beta_{1} =0 $, for Schwarzschild-AdS black hole, and $ l =0 $, $ \beta_{1} =0 $, for Schwarzschild black hole. Also, $ \alpha=0 $, $ \alpha=0.5 $ for (a) and (b), respectively.}
	\label{pic:staticN-C}
\end{figure}

Exploiting  relations (\ref{cor}), (\ref{T0}) and (\ref{C0}), the leading-order corrected entropy
due to thermal fluctuations is given by
\begin{eqnarray}
S={\cal S}_{0}-\alpha\log\left[ -\frac{4\beta_1l^2\pi^{1/2}{\cal S}_{0}^{3/2}-6{\cal S}_{0}^{2}+2l^2\pi{\cal S}_{0}}{l^2\pi+3{\cal S}_{0}}\right] -2\alpha \log\left[\frac{2\beta_1 l^2\pi^{1/2}{\cal S}_{0}^{1/2}-3{\cal S}_{0}+l^2\pi}{4\pi^{3/2}l^2{\cal S}_{0}^{1/2}} \right].
\end{eqnarray}
 Using standard relation  $\left(M=\int T_H dS\right)$,  the corrected mass is calculated as
\begin{eqnarray}\label{Mcor}
M&=&   \frac{l^2\pi^{1/2}\beta_1{\cal S}_{0}+ l^2\pi {\cal S}_{0}^{1/2}- {\cal S}_{0}^{3/2}}{2l^2\pi^{3/2}}+ \alpha\frac{l^2\pi+3{\cal S}_{0}}{2l^2\pi^{3/2}{\cal S}_{0}^{1/2}}+\frac{\sqrt{3}\alpha}{l\pi}\cot^{-1}\left(\frac{l\pi^{1/2}}{\sqrt{3{\cal S}_{0}}}\right)\nonumber\\
&-&\frac{\alpha\beta_1}{4\pi}[3\log {\cal S}_{0}-2\log (l^2\pi +3{\cal S}_{0})]-2\alpha \frac{2\beta_1 l^2\pi^{1/2}{\cal S}_{0}^{1/2}-3{\cal S}_{0}+l^2\pi}{4\pi^{3/2}l^2{\cal S}_{0}^{1/2}}.\label{M}
\end{eqnarray}
The corrected heat capacity, using relation  $\left(C=T_H\frac{\partial S}{\partial T_H}\right)$, is computed as
\begin{eqnarray}\label{Ccor}
C&=& -\frac{4\beta_1l^2\pi^{1/2}{\cal S}_{0}^{3/2}-6{\cal S}_{0}^2+2l^2\pi {\cal S}_{0}}{l^2\pi +3{\cal S}_{0}}-2\alpha\nonumber\\
&-&\alpha\frac{\left(\pi ^2 l^4+3\beta_1  \sqrt{\pi }  l^2 \sqrt{{\cal S}_{0}} \left(\pi  l^2+{\cal S}_{0}\right)-6 \pi  l^2 {\cal S}_{0}-9 S^2\right) \left( 2l^2  \beta_1  \sqrt{\pi } {\cal S}_{0}^{3/2}+\pi l^2  -3 {\cal S}_{0}\right)}{{\cal S}_{0} \left(\pi  l^2+3 {\cal S}_{0}\right) \left( 2 l^2 \beta_1 \sqrt{\pi } \sqrt{{\cal S}_{0}}+\pi l^2 -3 {\cal S}_{0}\right)^2}.
\end{eqnarray}
Again, Eqs. (\ref{Mcor}) and (\ref{Ccor}) for $ \alpha=0 $, and $ \beta_1=0 $, convert to Eqs. (\ref{M-sads}) and (\ref{C-sads}), and for $ \alpha=0 $, $ \beta_{1}=0$ and $\Lambda=0 $, convert to Eqs. (\ref{M-s}) and (\ref{C-s}),
which represent the Schwarzschild-AdS and Schwarzschild black holes, respectively.
These thermodynamic parameters are plotted in terms of horizon radius $r_{+}$ (see Figs.  
\ref{pic:static-M}, \ref{pic:staticN-M}, \ref{pic:static-C} and \ref{pic:staticN-C}).

From Fig.~\ref{pic:static-M}(a), (for $ \alpha=0$), it can be seen that, 
mass of the black hole has a maximum value
at $r_{+}=r_{m}=2.31$, and it become zero at two points, $r_{+}=r_{01}=0$ and $r_{+}=r_{02}=4$.
Also, as can be seen from Fig.~\ref{pic:static-M}(b) (for $ \alpha>0$), by increasing $ \alpha $, 
the number of zero points of mass are not changing, but they shift towards larger values.

Moreover, Fig. \ref{pic:staticN-M} shows the comparison of variation of mass versus horizon radius for a static black hole in $ f(R) $ gravity and  the Schwarzschild-AdS and Schwarzschild black holes of standard general relativity.
As one can see, for both cases ($ \alpha=0 $ and $ \alpha=0.5 $), mass variations are somewhat the same for a static black hole in $ f(R) $ gravity and the Schwarzschild-AdS black hole. But for the Schwarzschild black hole the mass variation is quite different. 

In addition, plot of the heat capacity  can be seen in 
Fig.~\ref{pic:static-C}. In this case, we find that, for the case $ \alpha=0$ (Fig.~\ref{pic:staticN-C}(a)),
for $0<r_{+}<r_{m} $, the heat capacity is in the negative region (unstable phase), 
then, at  $ r_{+}=r_{m} $, it takes type one phase transition, in which $ C(r_{+}=r_{m})=0) $, after that, for  $ r_{+}>r_{m} $, it
becomes positive valued (stable).
So, for the case $ \alpha=0$, this black hole has a type one phase transition.
However, (for $ \alpha>0$), as can be seen from Fig.~\ref{pic:static-C}(b), by increasing $ \alpha $,
the number of zero points of the heat capacity are changing to the three zero points.
In   other words, for the case $ \alpha=0 $, in addition to one zero  two other zeroes also appear.

Also, Fig. \ref{pic:staticN-C} shows the comparison of variation of the heat capacity versus horizon radius for a static black hole in $ f(R) $ gravity and  the Schwarzschild-AdS and Schwarzschild black holes of standard general relativity.
As we can be seen, for the case $ \alpha=0 $, the heat capacity variations are somewhat the same for a static black hole in $ f(R) $ gravity and the Schwarzschild-AdS black hole. But for the Schwarzschild black hole the heat capacity variation is quite different and  it is always in the negative phase and it is unstable. 
Moreover, for the case $ \alpha>0$, these  comparisons are somewhat different. For this case, all three black holes have completely different heat capacity variations. the Schwarzschild black hole has no zero and is therefore always unstable. The Schwarzschild-AdS black hole has a zero and therefore, as mentioned earlier, has a phase transition of a type one. The static black hole in $ f(R) $ gravity (for $ \alpha>0$), has more zeros as alpha increases and therefore has more phase transitions.  

\subsection{Thermodynamic geometry}
\begin{figure}[h]
	\centering	
		\includegraphics[width=0.4\textwidth]{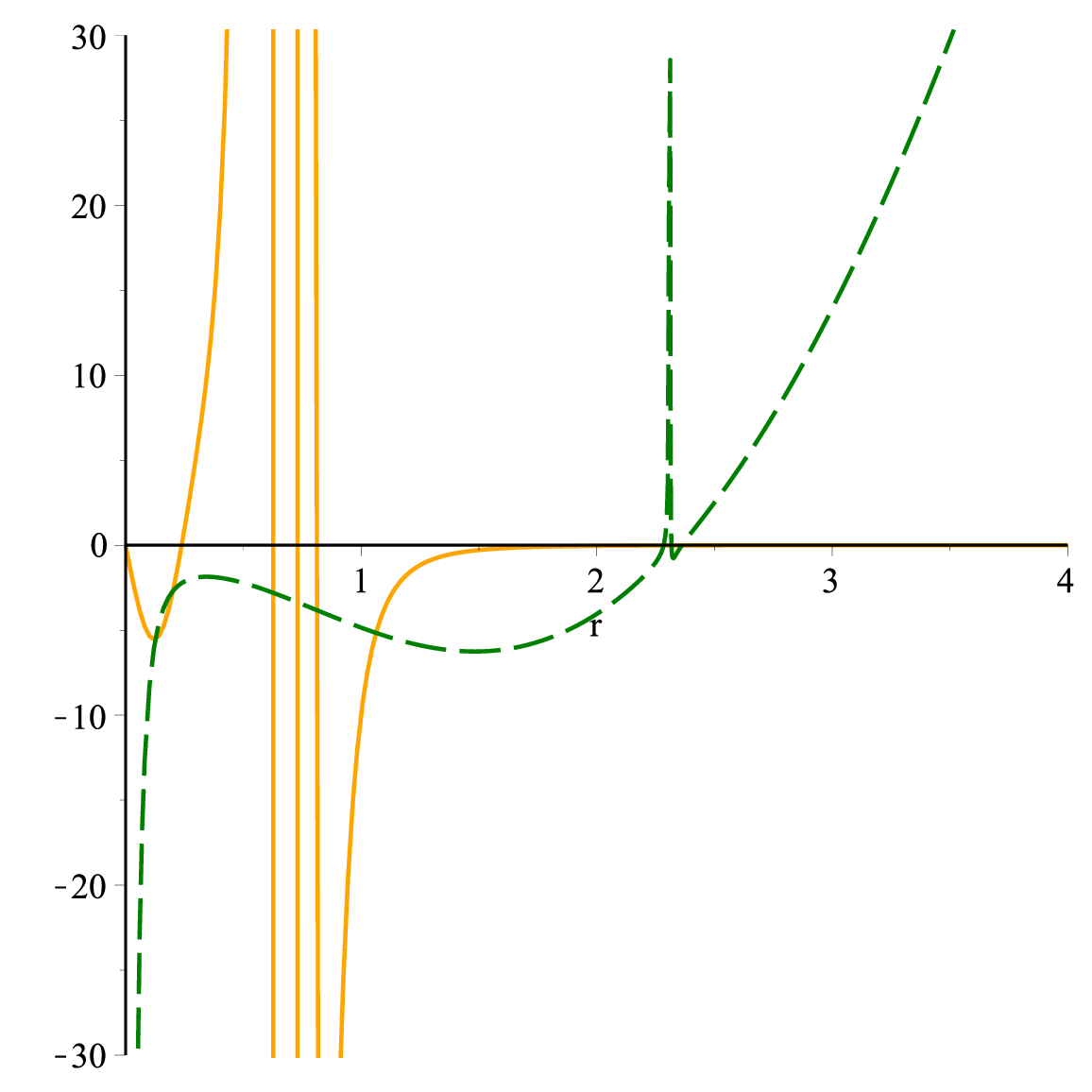}	
	\caption{Curvature scalar variation of Weinhold metric (orange continuous
line) and the heat capacity of a static black hole (green dash line) in
terms of the radius of the horizon r+, for $ \alpha=0.5 $, $ l =4.0 $, $ \beta_{1} =10^{-4} $.}
 \label{pic:staticWein}
\end{figure}

In this section, using the geometric technique of Weinhold, Ruppeiner and GTD
metrics of the thermal system, we construct the geometric structure 
for a static black hole in $ f(R)  $ gravity. 
In this case, the extensive variables are,
$N^{r}=(l, \beta_{1})$. The
Weinhold metric is defined as the Hessian in
the mass representation as follows~\cite{Weinhold}
 \begin{equation}\label{Weinhold}
 g^{W}_{ij}=\partial_{i}\partial_{j}M(S,N^{r}).
 \end{equation}
We can write the Weinhold metric for this system as follows
 \begin{equation}
 g^{W}_{i j}=\partial _{i}\partial _{j}M(S,l, \beta_{1}).
 \end{equation}
The line element corresponding to Weinhold  metric is given by 
\begin{eqnarray}
ds^{2}_{W}=M_{SS}dS^{2}+M_{l l}dl^{2}+M_{\beta_{1}\beta_{1}}d\beta_{1}^{2}\nonumber\\
  2M_{Sl}dSdl +2M_{S\beta_{1}}dSd\beta_{1} +2M_{l\beta_{1}}dl d\beta_{1} ,
\end{eqnarray}
 therefore
\begin{equation}
g^{W}=\begin{bmatrix}
M_{SS} & M_{Sl} & M_{S\beta_{1}}\\
M_{l S} & M_{l l} &M_{l \beta_{1}}\\
M_{\beta_{1} S} & M_{\beta_{1} l} & 0
\end{bmatrix}.
\end{equation}
The components of above matrix can be found using the expression of
$M$, given in Eq.~(\ref{M}). 
Since the equation of the curvature
scalar of the Weinhold metric is too
large, so we demonstrate it in Fig.~(\ref{pic:staticWein}).

\begin{figure}[h]
	\centering
	 \subfigure[$ \alpha=0.5 $]{
		\includegraphics[width=0.4\textwidth]{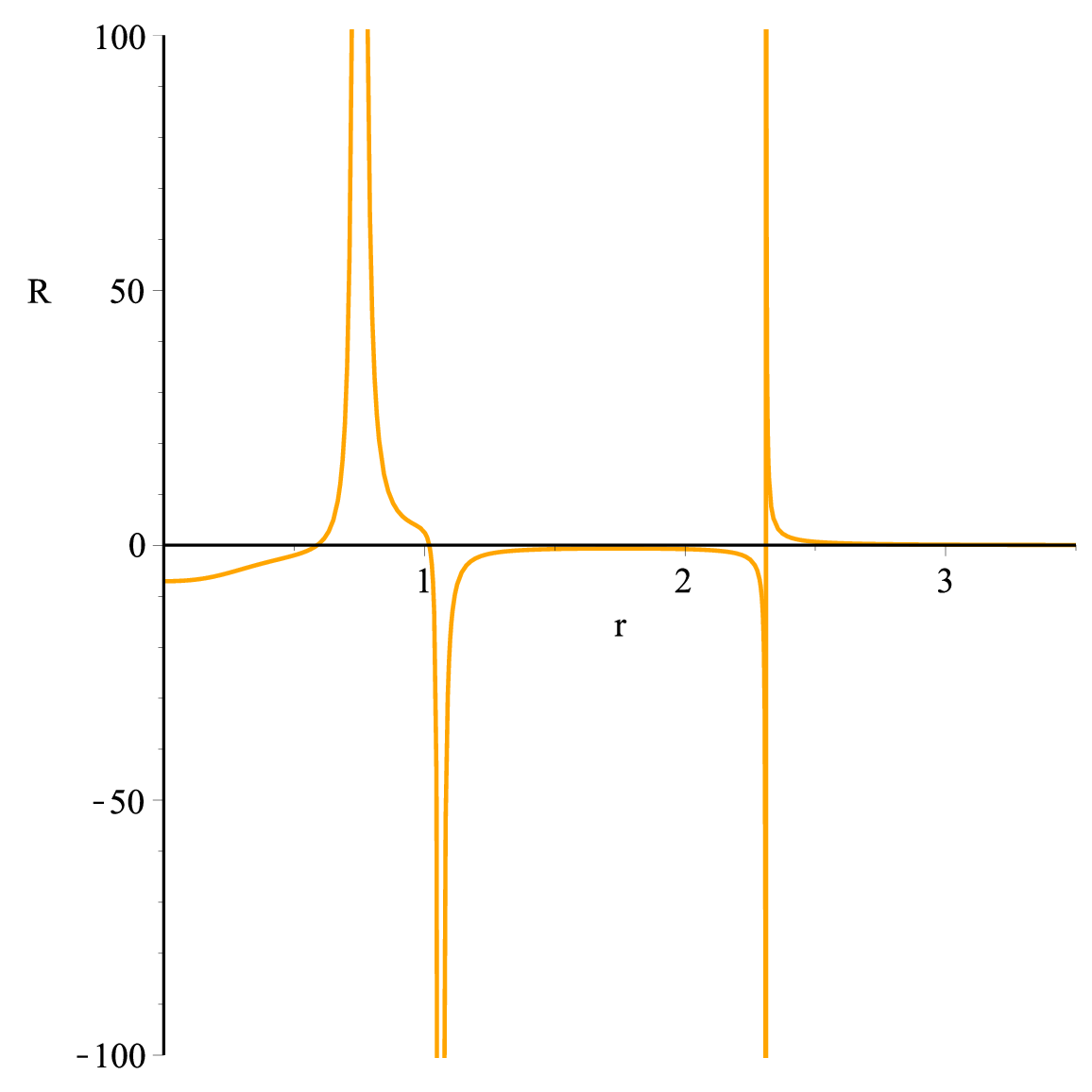}
	}
	\subfigure[$ \alpha=0 $]{
		\includegraphics[width=0.4\textwidth]{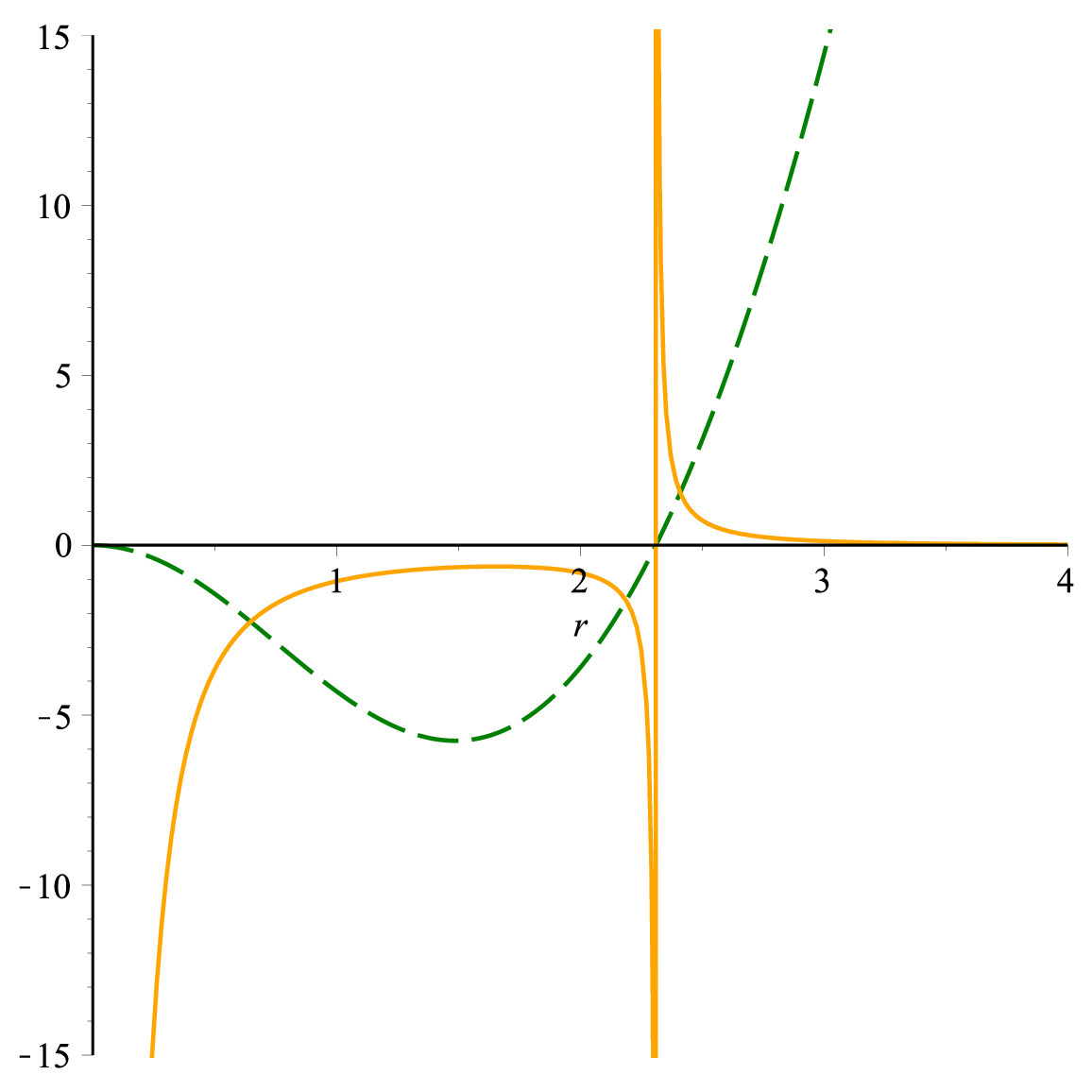}
	}
     \subfigure[$ \alpha=0.5 $]{
		\includegraphics[width=0.4\textwidth]{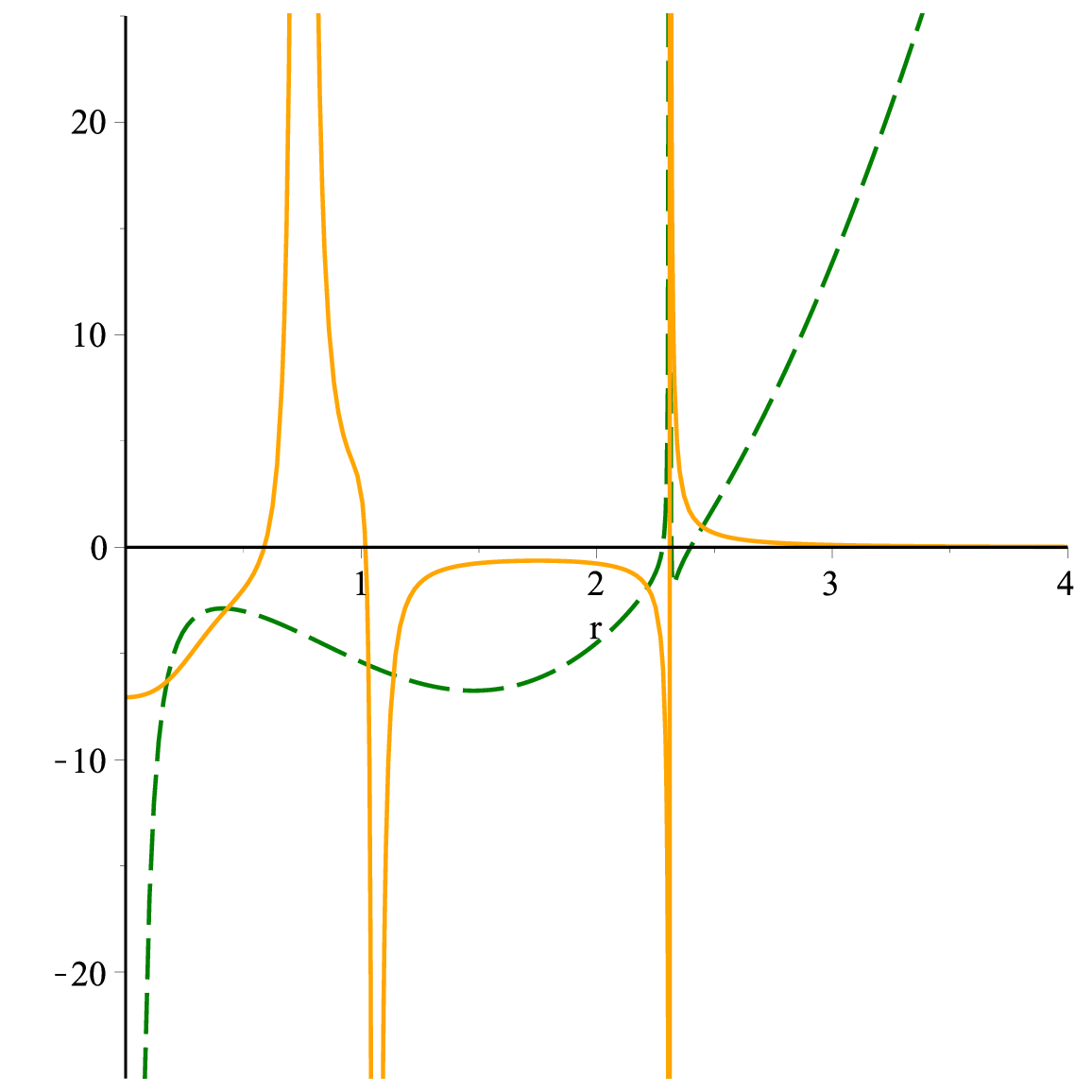}
	}
     \subfigure[Closeup of figure (c)]{
		\includegraphics[width=0.4\textwidth]{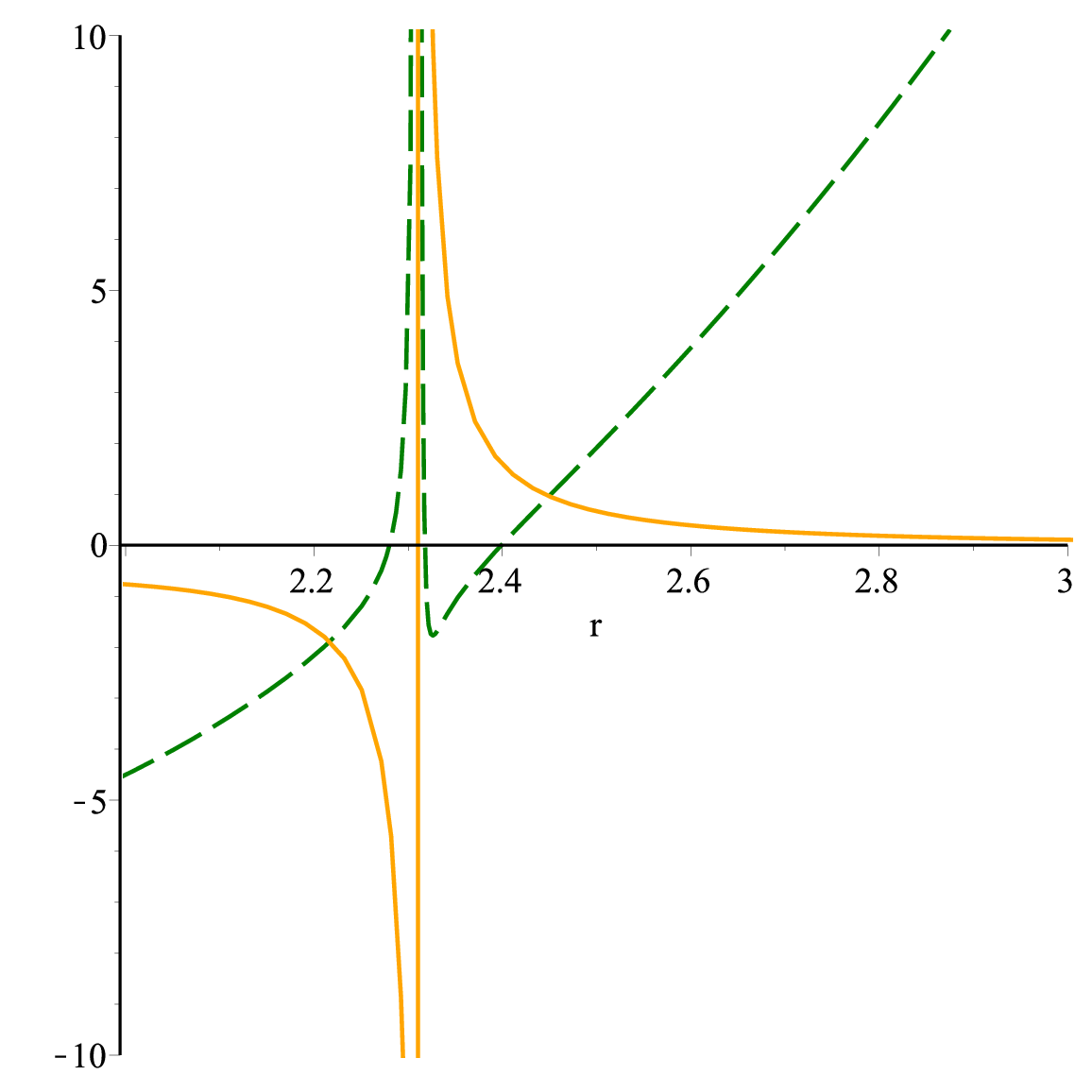}
	}
	\caption{Curvature scalar variation of Ruppeiner metric (orange continuous line) and 
the heat capacity of a static black hole (green dash line) in terms
of $ r_{+} $ for $ l=4.0 $, $ \beta_{1} =10^{-4} $.}
 \label{pic:staticRup}
\end{figure}

It can be observed from Fig.~\ref{pic:staticWein} that the singular points of curvature scalar of 
Weinhold metric do not coincide with zero points of heat capacity, so, in this case we can't 
find any physical information about the system from the Weinhold method.
In following, we use
Ruppeiner method, which is conformaly transformed to the Weinhold
metric. The Ruppeiner metric is defined by~\cite{Ruppeiner,Salamon,Mrugala:1984}
\begin{equation}
ds^{2}_{R}=\frac{1}{T}ds^{2}_{W}.
\end{equation}
The   matrix corresponding to the metric components of Ruppeiner method  is as following:
\begin{equation}
g^{R}=\left(\frac{4l^{2}\pi^{\frac{3}{2}}S_{0}^{\frac{1}{2}}}{2l^{2}\pi^{\frac{1}{2}}S_{0}^{\frac{1}{2}}\beta_{1} -3S_{0}+l^{2}\pi}\right)\begin{bmatrix}
M_{SS} & M_{Sl} & M_{S\beta_{1}}\\
M_{l S} & M_{l l} &M_{l \beta_{1}}\\
M_{\beta_{1} S} & M_{\beta_{1} l} & 0
\end{bmatrix}.
\end{equation}
Plot of the scalar curvature of the Ruppeiner metric is shown in Fig.~\ref{pic:staticRup}(a).
Also, plots of the curvature scalar of the Ruppeiner metric and the heat capacity, in
terms of  $ r_{+} $, are demonstrated in Fig.~\ref{pic:staticRup}(b), (c) and (d). 
As one  can see from Fig.~\ref{pic:staticRup}(b), for the case $ \alpha=0 $, 
the heat capacity has only one zero (the phase transition point), however, the singular point of the curvature scalar 
of Ruppeiner metric  completely coincides with zero point of the heat capacity.
On the other hand, for the case $ \alpha>0 $ (see Fig.~\ref{pic:staticRup}(c), (d)), the heat capacity has three zero points, however, the singular points of the curvature scalar of Ruppeiner metric are not completely coinciding with the
zero points of the heat capacity of the static black hole. In fact, only one 
of the singular point of the curvature scalar of Ruppeiner metric is completely coinciding with one 
of the zero point of the heat capacity. In the other word, by increasing $ \alpha $, this adaptation is reduced. 

\begin{figure}[h]
	\centering	
		\includegraphics[width=0.4\textwidth]{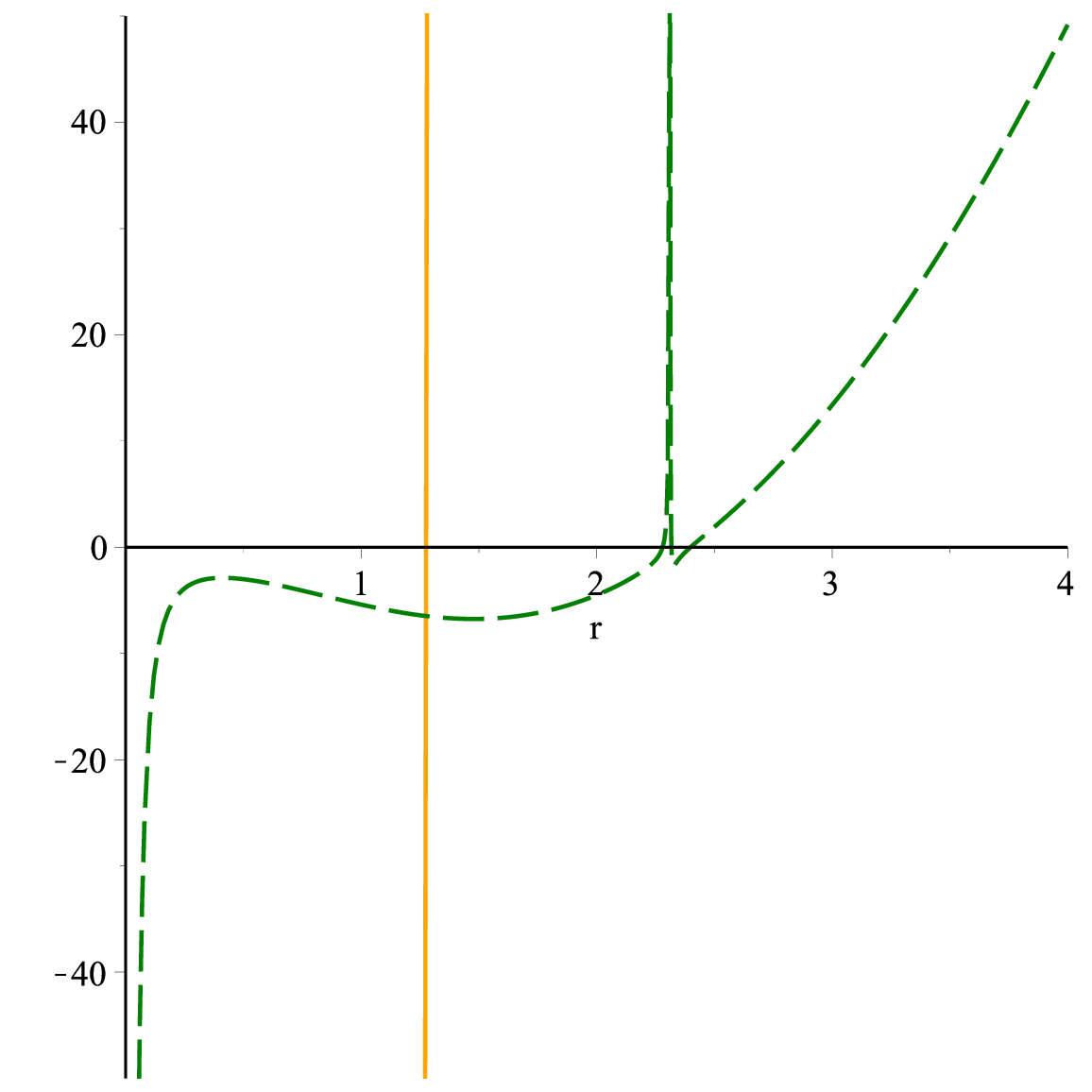}	
	\caption{Curvature scalar variation of GTD metric (orange continuous
line) and the heat capacity of a static black hole (green dash line) in
terms of the radius of the horizon r+, for $ \alpha=0.5 $, $ l =4.0 $, $ \beta_{1} =10^{-4} $.}
 \label{pic:staticGTD}
\end{figure}

In the next step, we will investigate the GTD  method.
The general form of the metric in   GTD  
method is as following~\cite{Quevedo}:
\begin{equation}\label{GTD}
g=\left(E^{c}\frac{\partial\Phi}{\partial E^{c}}\right)\left(\eta_{ab}\delta^{bc}\frac{\partial^{2}\Phi}{\partial E^{c}\partial E^{d}}dE^{a}dE^{d}\right),
\end{equation}
where 
\begin{equation}
\frac{\partial\Phi}{\partial E^{c}}=\delta_{cb}I^{b},
\end{equation}
in which $ \Phi $ is the thermodynamic potential, $ I^{b} $  and $ E^{a} $ are the intensive and extensive 
thermodynamic variables.
So,  according to Eq.~(\ref{GTD}), the metric for this thermodynamic system is as following:
\begin{equation}
g^{GTD}=(SM_{S}+l M_{l}+\beta M_{\beta_{1}})\begin{bmatrix}
-M_{SS} & 0 & 0\\
0 & M_{ll} & M_{l \beta_{1}}\\
0 & M_{\beta_{1} l} & 0
\end{bmatrix}.
\end{equation}
The plot of the scalar curvature of   GTD   metric is shown in Fig.~(\ref{pic:staticGTD}).
It can be observed from Fig.~\ref{pic:staticGTD}, the singular point of curvature scalar of GTD 
metric does not coincide with zero point of heat capacity, so, in this case we can't find any physical 
information about the system from the GTD method.


\section{Conclusions}\label{section4}

In this paper, we studied small statistical fluctuations around equilibrium 
for a static black hole in $f(R)$ gravity and analyzed thermodynamic quantities 
of this black hole according to corrected thermodynamic entropy.  We have found  that the Hawking temperature is a decreasing 
function of horizon radius.  We  have utilized the
 expressions of Hawking temperature and uncorrected specific heat  to evaluate the correction to
 the entropy of a static black hole in $f(R)$ gravity due to   thermal fluctuation. Also, we 
 have computed the first-order corrected mass and heat capacity by utilizing  the standard thermodynamical definitions.
From graphical study, we observed that the uncorrected 
mass of the black hole gets a maximum value
at $r_{+}=r_{m}=2.31$, and takes zero value at two points, $r_{+}=0$ and $r_{+}=4$. 
However, for the corrected mass  with positive correction coefficient $ \alpha $, 
the number of zero points of mass do not changing, but  they appear at larger horizon radius.
For the case of heat capacity, we have found that  
 the uncorrected heat capacity has negative values (unstable phase) for $0<r_{+}<r_{m} $  and type one phase transition  takes place for black holes undergo  at  $ r_{+}=r_{m} $  as $ C(r_{+}=r_{m})=0 $. For  $ r_{+}>r_{m} $,  the heat capacity
takes positive value  (stable).
So, one may conclude that without considering thermal fluctuation, this black hole has a type one phase transition.
However, due to e thermal fluctuation,
the number of zero points of the heat capacity changes to the three zero points.

Also, we investigated the thermodynamic geometry of this black hole and 
plotted thermodynamic quantities and curvature scalar of Weinhold, Ruppeiner 
and GTD methods in terms of horizon radius $ r_{+} $.
We observed that, the singular points of curvature scalar of Weinhold and GTD methods, 
for both $ \alpha=0 $ and $ \alpha>0 $ cases, are not coincided with zero point of 
heat capacity (the phase transition points). This suggests that we could not find any physical information 
about the system from these two methods. 
However, for the case $ \alpha=0 $, we realized that the heat capacity has only one zero, 
and the singular point of the curvature scalar of Ruppeiner metric is completely coinciding 
with it, moreover for the case $ \alpha>0 $, the heat capacity has 
three zero points, and the singular points of the curvature scalar of Ruppeiner metric are 
not completely coinciding with zero points of the heat capacity of this black hole and only one 
of the singular point of the curvature scalar of Ruppeiner metric is completely coincide with one 
of the zero point of the heat capacity, that's mean by increasing $ \alpha $, this adaptation is reduced.

\end{document}